# A SIMULINK CIRCUIT MODEL FOR MEASUREMENT OF CONSUMPTION OF ELECTRIC ENERGY USING FREQUENCY METHOD

by

Mr. Emmanouil N. Markoulakis

Nov 2018

(Beng. MSc. Electronic and VLSI Systems Engineering UMIST. U.K., MSc. Technical University of Crete), (Technological Educational Institute of Crete, Chania 73133, GREECE), markoul@chania.teicrete.gr )

Prof. Dr. Emmanuel N. Antonidakis

(Technological Educational Institute of Crete, Chania 73133, GREECE, ena@chania.teicrete.gr )

Prof. Dr. George S. Stavrakakis

(Technical University of Crete, Chania 73100, GREECE, gstavr@electronics.tuc.gr )

**ABSTRACT**

The following analysis and presentation summarises the **implementation** in **Matlab / Simulink** *environment* of a prototype model digital energy circuit for measurement of the consumption of electric energy of an electrification network in the installations of customer of electric energy that can be applied as part of any consumption **smart meter**.

## 1. Introduction

The method of direct calculation of the consumed electric energy as the product of *electric power* over *time* cannot be used exclusively in a permanently installed, interminable electric energy meter at the customer premises, because of additive, accumulated innate fault as a result of the before mentioned method of continuous floating point calculations and approximation error increasing with the time of consumption [1][2].

An alternative method of electric energy measurement should be sought and applied in this case which will give precise, and reliable measurements correlated with an error independent from the consumption time.

In the following pages we propose, analyze and simulate a system blog diagram of such an energy circuit which uses a measurement method based on the production of energy pulses with a frequency that corresponds directly to the consumed electric energy every time. The simulation results together with graphic representations are demonstrated and prove the precision, stability and reliability of the proposed frequency method.

## 2. Analysis

The simulated energy circuit is depicted in Figure 1. We will not refer here in detail on the parametric design of the circuit in **Matlab / Simulink** environment. More specific information on the design can be found in the simulation computer file which was uploaded on the internet and which is available at the following addresses:

https://drive.google.com/file/d/0B0A8uTBvEiQRcGJlbGlOMjhQVFU/view?usp=sharingas .

The time of simulation is given in seconds and the simulation model that was created, can simulate any combination of electric power value and time and measure the electric energy that was consumed in KWh units.

A maximum current amplitude accommodated at the customer's installation in which the energy circuit is to be applied, is assumed to be 40A RMS and *with a maximum allowed* single phase Voltage value at 240V RMS. Provided that these limits are exceeded the meter "cuts" the electricity on the customer's installation for reasons of protection. The **i** and **v** shown in Figure 1 are the alternating ac measurement signals of current and voltage in the installation (in the simulator their peak values are given), that are provided by the corresponding sensors of current and voltage included in the meter (current and voltage transducers). The signal attenuators **Gain** and **Gain 1** shown, change the level of these signals in levels suitable for digitization (0-5V). Accordingly, with the help of a **digital multiplier (Product)** and the attenuation signal function unit of a factor **1/√2 (u/sqrt 2),** two different power measurement signals are derived from, the **instantaneous power signal**, and its *active* value counterpart thus, **Instantaneous (Active) Power Signal.** The last signal can be led directly to a calculating unit (e.g. microcontroller), where its positive peak value is selected each time and then multiplied continuously with consumption time in order to calculate the consumed energy.

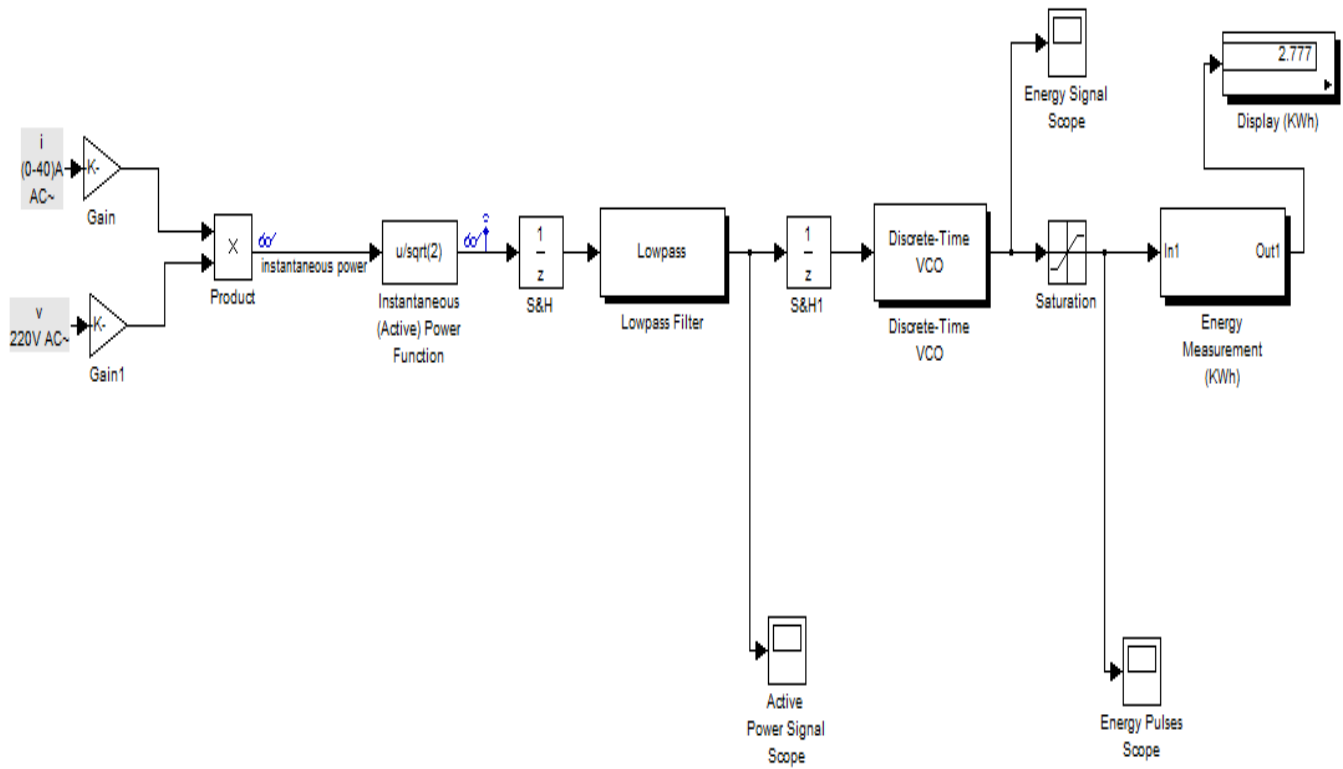

Figure 1. The Energy Circuit

However, this method as we mentioned before accumulates continuously with time, error of approximation from the continuous floating point calculations and is therefore judged as inadequate for our aim. Alternative, a frequency method is implemented for finding the consumed electric energy. Thus if we receive the full - scale signals of i and v of the meter's specifications which are 40A RMS and 240V RMS accordingly, this equals an amount of 9.6 KW RMS power consumption and the full - scale simulated signals of Figures 2&3 are taken on the corresponding points of our energy circuit. The signal then from Figure 3 is lead via sampling unit S and H (sample and hold) into the entry of a digital **Low pass filter** that exports each time the **average value** of its input

.

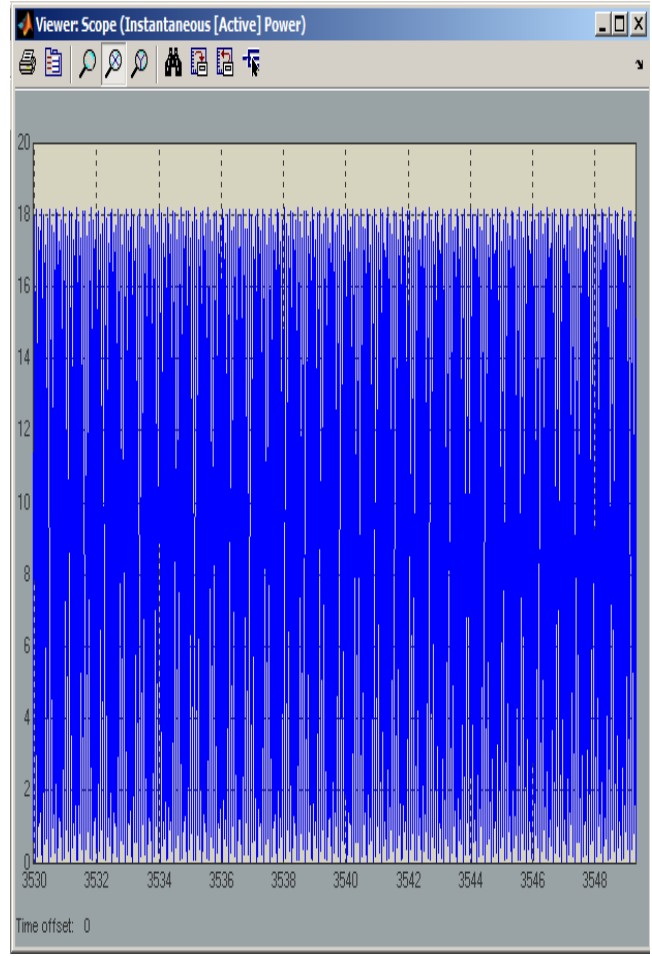

Figure 3. Instantaneous (Active) Power

In that way, each measured value of electric power entered in the energy circuit is correlated with a unique **average** value of *voltage* in the exit of the filter (**Active Power Signal**). In our case in question, in full - scale operation of the energy circuit with 9.6 KW (40A, 240V) entry power, the exit of the filter has been regulated to be 9.6V *(shown in Figure 4).*

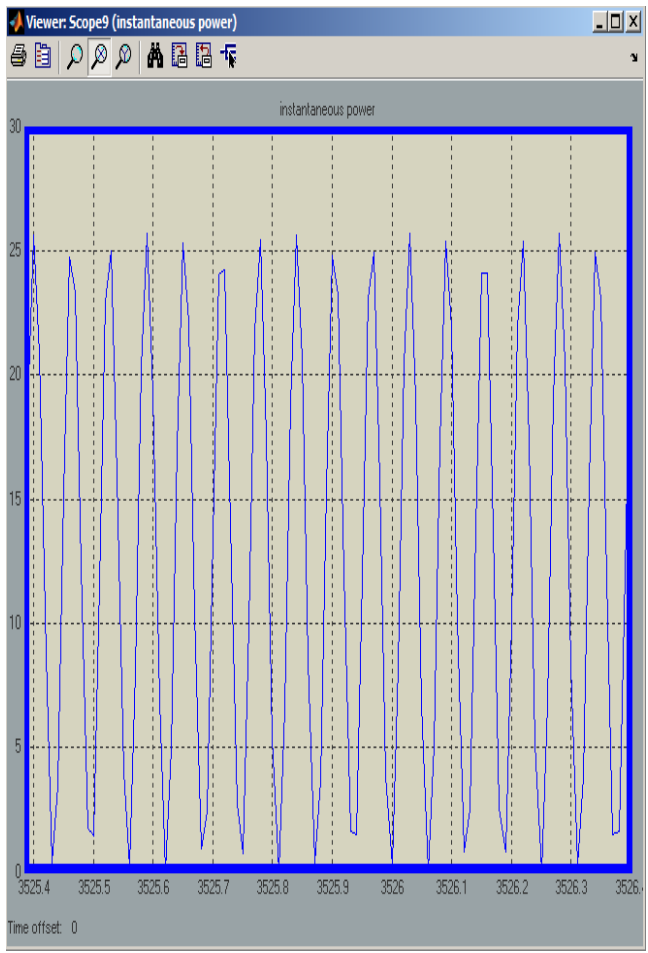

Figure 2. Instantaneous Power

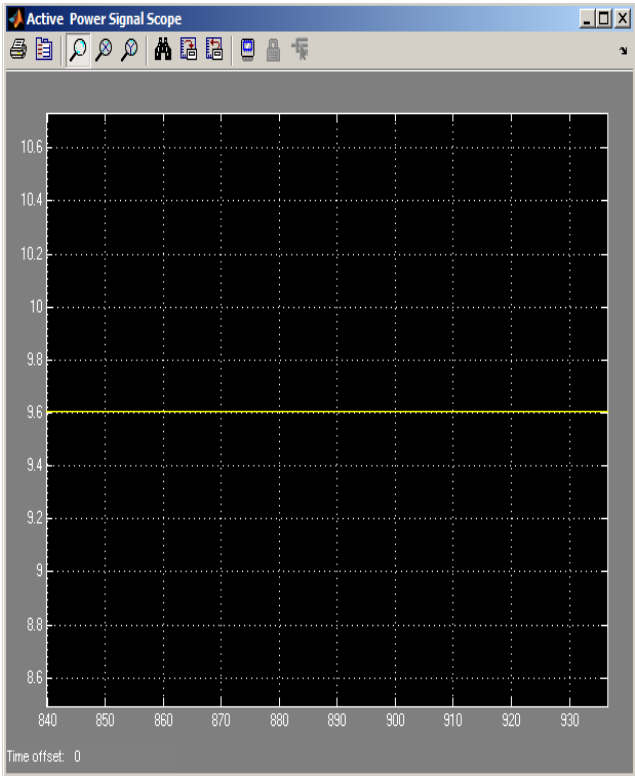

Figure 4. Active Power Signal

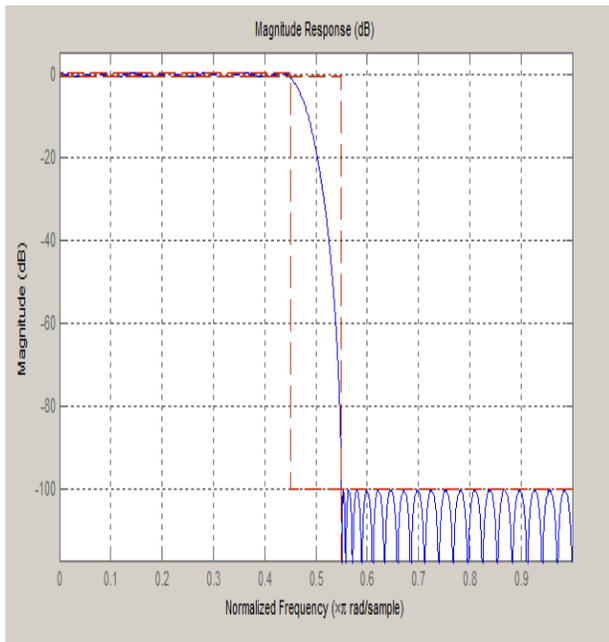

Figure 5. Filter Response

Following, this average output value of the filter drives a **digital sine wave oscillator**, the oscillating frequency of which is controlled by the voltage value of the output signal of the filter *(i.e. VCO, Voltage Controlled Oscillator)*.

The sensitivity of the VCO, **Hz / V**, has been regulated so that an entry power in the energy circuit of **1 KW** produces a signal of 1000 sine cycles per hour (3600 sec) which directly correlates with a consumption of electric energy of **1 KWh.** Finally, the sine wave signal of the oscillator **(Energy Signal)**, drives an electronic switch circuit **(Saturation)** after its negative semi – period is first eliminated, for the production of energy square pulses **(Energy Pulses, 50:50 duty cycle, 5V Amplitude)** used for the measurement of the consumed electric energy in the installation. The proportion here again is **1000 pulses/KWh**.

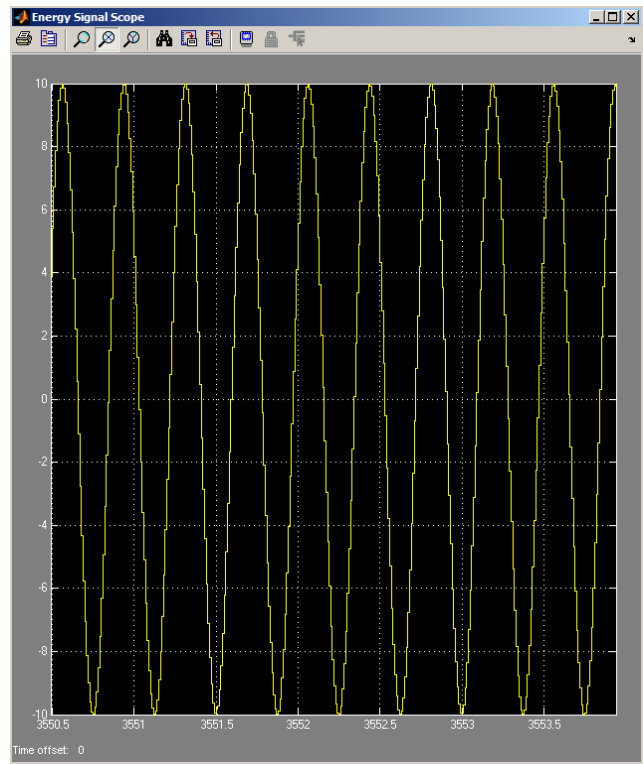

Figure 6. Energy Signal

These energy pulses can then be led directly into a microcontroller part of a smart meter or other remote calculating system for enumeration of the pulses and the calculation of consumed electric energy in KWh units.

The proposed however energy circuit demonstrates complete autonomy and exports its measurement of energy in units **KWh** via a **LCD display** with coding *BCD to 7 segment* and precision three decimal digits (e.g. 23.589 KWh).

The Energy Measurement subsystem contains a triggered 32-bit BCD *counter* and a digital *divider* unit (**u / 1000**) for the transformation in KWh.

## 3. Results

Various scenarios of simulation were executed in order to prove the precision stability and reliability of the proposed energy circuit.

The results of simulation are presented analytically in this section with the form of tables and graphic representations that show percentage of error (± of error %) of the Simulated Energy Value compared to the expected theoretical value, Energy Value Expected, for various combinations of consumption time and entry power.

Thus we have simulations for consumption times ranging from one minute to one week and power from 22W (100mA, 220V) up to the biggest permissible by the meter, 9.6KW (40A, 240V).

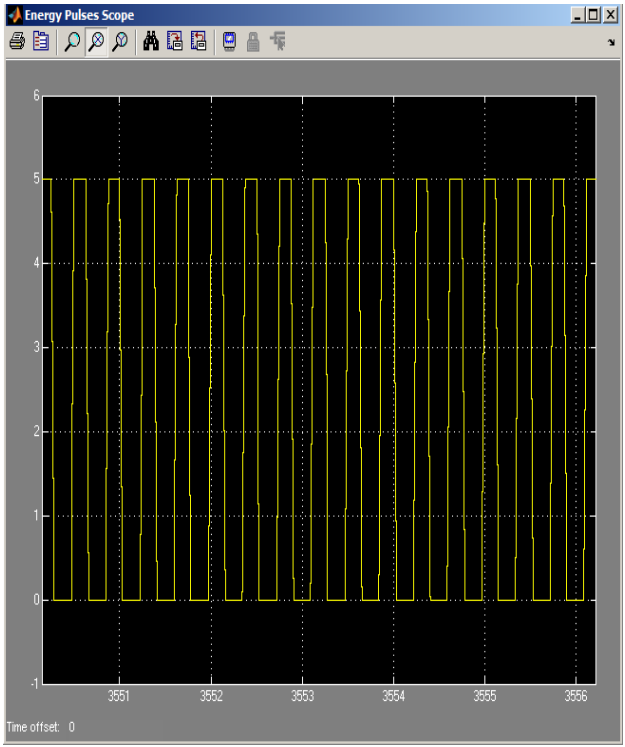

Figure 7. Energy Pulses

The simulation results are summarised as following:

Thus, the Energy Pulses are imported in the energy calculation unit (Energy Measurement) for the calculation of energy in KWh units.

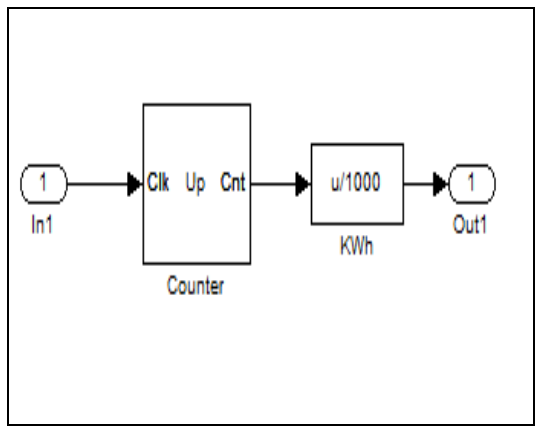

Figure 8. Energy measurement subsystem

**Scenario 1**: ±Error % vs Time   I=40A, V=240V, P=9.6KW

| Consumption Time (sec) | hours | Energy Expected (KWh) | Energy Simulated (KWh) | ±error % |
|---|---|---|---|---|
| 60 | 0.0166 | 0.16 | 0.158 | **-1.25** |
| 1800 | 0.5 | 4.8 | 4.793 | **-0.14** |
| 3600 | 1 | 9.6 | 9.592 | **-0.083** |
| 7200 | 2 | 19.2 | 19.192 | **-0.041** |
| 10800 | 3 | 28.8 | 28.798 | **-0.038** |
| 14400 | 4 | 38.4 | 38.379 | **-0.053** |
| 21600 | 6 | 57.6 | 57.567 | **-0.057** |
| 28800 | 8 | 76.8 | 76.76 | **-0.052** |
| 43200 | 12 | 115.2 | 115.14 | **-0.052** |
| 86400 | 24 | 230.4 | 230.268 | **-0.057** |

**Scenario 2**: ±Error % vs Power  V=220V, Consumption Time T=3600s (1hour)

| Current I (A) | Energy Expected (KWh) | Energy Simulated (KWh) | ±error % |
|---|---|---|---|
| 0.1 | 0.022 | 0.022 | **0** |
| 1 | 0.22 | 0.22 | **0** |
| 2 | 0.44 | 0.439 | **-0.227** |
| 5 | 1.1 | 1.099 | **-0.09** |
| 7 | 1.54 | 1.539 | **-0.06** |
| 10 | 2.2 | 2.198 | **-0.09** |
| 15 | 3.3 | 3.297 | **-0.09** |
| 20 | 4.4 | 4.397 | **-0.068** |
| 25 | 5.5 | 5.495 | **-0.09** |
| 30 | 6.6 | 6.595 | **-0.075** |
| 35 | 7.7 | 7.694 | **-0.078** |
| 40 | 8.8 | 8.793 | **-0.079** |

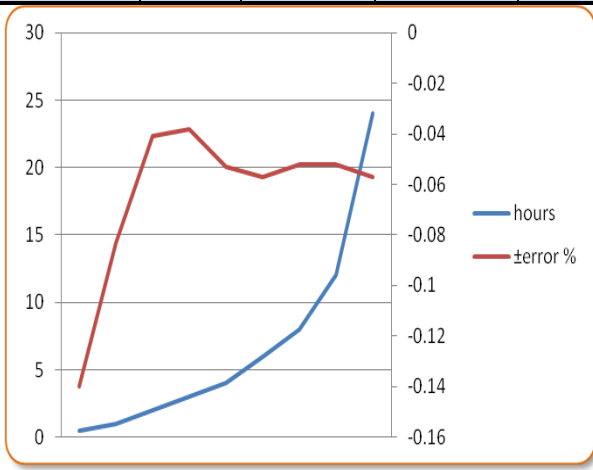

Table 1.  ±Error% vs Time

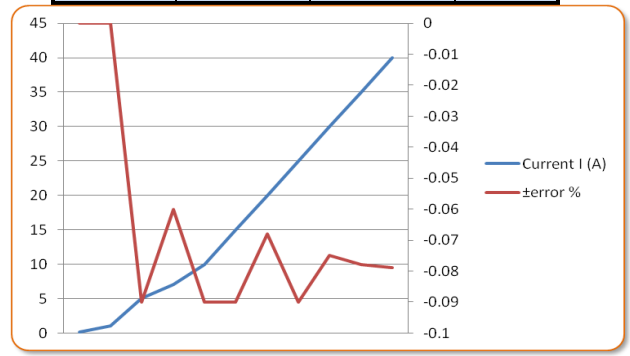

Table 2.  ±Error% vs Power

**Scenario 3**: ±Error % vs Power vs Time(long)

**V=220V, Consumption Time(long) T=604800 sec (1 Week)**

| Current I (A) | Energy Expected (KWh) | Energy Simulated (KWh) | ±error % |
|---|---|---|---|
| 1 | 36.96 | 36.939 | **-0.056** |
| 20 | 739.2 | 738.779 | **-0.056** |
| 40 | 1478.4 | 1477.558 | **-0.056** |

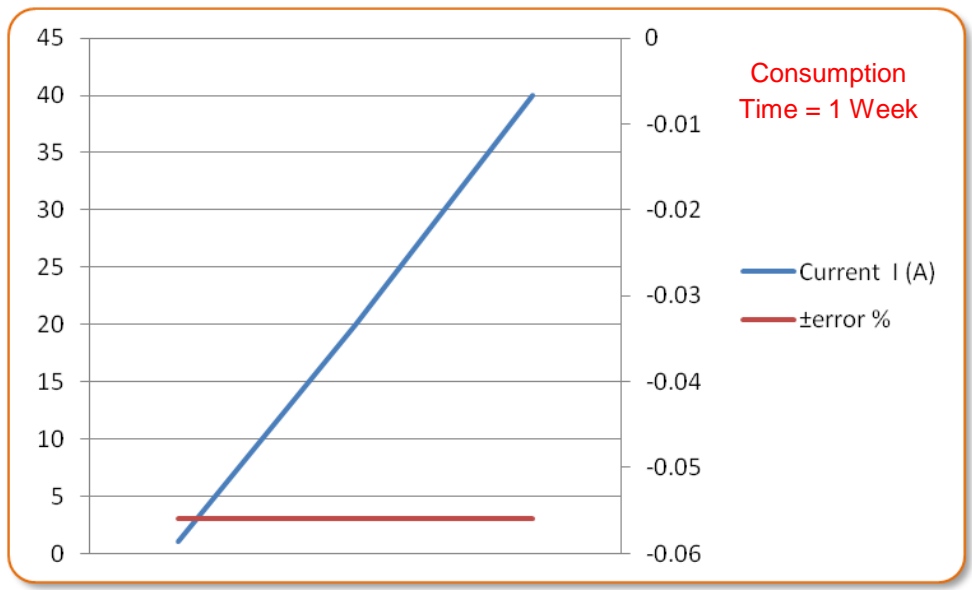

Consumption Time = 1 Week

Table 3. ±Error% vs Power vs Time(long)

## 4. Conclusions

The results of simulation showed generally a uniform very small *error* in the order of **-0.05% ÷ -0.09%** in all the span of possible electric loads and power applied to the meter as well as time periods of measurement that vary in the particular simulation, from 60 seconds to one week.

Therefore, the energy circuit and method of measurement that we propose, demonstrates as it was proven big precision, stability and reliability and most important as it was also proven, that *it does not accumulate error* in measurement over *time*. On the contrary, the already **very small** error of the meter *presents bending tendency with the increase of time periods of measurement of the consumed electric energy* and tends to become constant independent from electric power in the installation (shown in Table 3, simulation for one week).

Practically, our energy circuit has *null fault* for relative large measurement periods such as 1 Month, 2 Months… 4 Months which is actually the norm for a permanently installed energy consumption meter on the customer of electric energy premises, connected to an electrification network, since the billing service company makes measurements and it usually publishes accounts per intervals of no less than two month periods.

## 5. Epilogue

It was proposed and implemented with success in a simulator, a model of an energy circuit for measurement of single phase electric energy consumption from an electrification network connected to the installations of a customer of electric energy. With direct application in all Energy Smart Meters.

The capability and advantages of the proposed and implemented frequency measurement method were proved.

*"The minimisation of fault of measurement of energy in this method against the conventional calculating method lies in the make that the measurement of energy value however many floating decimal points it consists of (e.g. 3.296 KWh), is always correlated to an **integer** number of energy pulses used for the measurement ".*